\documentclass[11pt]{article}
\usepackage{moriond,epsfig}

\def\be{\begin{equation}}
\def\ee{\end{equation}}
\def\bea{\begin{eqnarray}}
\def\eea{\end{eqnarray}}

\usepackage{xspace}
\usepackage{lineno}

\begin{document}

\newcommand{\dzero}     {D\O\xspace}
\newcommand{\wplus}     {$W+$jets\xspace}
\newcommand{\zplus}     {$Z+$jets\xspace}
\newcommand{\muplus}    {$\mu +$jets\xspace}
\newcommand{\eplus}     {$e +$jets\xspace}
\newcommand{\ljets}     {$\ell +$jets\xspace}
\newcommand{\ttbar}     {$t\bar{t}$\xspace}
\newcommand{\met}       {$\not\!\!E_T$\xspace}


\vspace*{4cm}
\title{Searches for New Physics in Top Events at the Tevatron}

\author{ A.W. Jung for the CDF and \dzero collaboration \\
FERMILAB-CONF-12-141-PPD}

\address{Fermi National Accelerator Laboratory,\\
P.O. Box 500, Batavia, IL, 60510, USA}

\maketitle\abstracts{Recent results of searches for new physics in top events at the Tevatron are presented. In case of CDF three searches are discussed using $6.0$ to $8.7~\mathrm{fb^{-1}}$ of data, with the latter being the final CDF data sample available for this kind of analysis. CDF carried out a search for Top + jet resonance production, dark matter production in association with single top and boosted tops. No signs of new physics are observed and instead upper limits are derived. \dzero used $5.3~\mathrm{fb^{-1}}$ of data and searched for a narrow resonance in \ttbar production and a time dependent \ttbar cross section, which would reveal a violation of Lorentz invariance. However, no signs for deviations from Standard Model are seen and instead upper limits for non-Standard Model contributions are calculated.}

\section{Introduction}
The $top$ quark is the heaviest known elementary particle and was discovered at the Tevatron $p\bar{p}$ collider in 1995 by the CDF and \dzero collaboration \cite{top_disc1,top_disc2} with a mass around $173~\mathrm{GeV}$. The production is dominated by the $q\bar{q}$ annihilation process with 85\% as opposed to gluon-gluon fusion which contributes only 15\%. The measurements presented here are performed using either the all-jets final state or the \ljets~channel. Within the \ljets~final state one of the $W$ bosons (stemming from the decay of the $top$ quarks) decays leptonically, the other $W$ boson decays hadronically. For the all-jets final state both $W$ bosons decay hadronically. The branching fraction for top quarks decaying into $Wb$ is almost 100\%. Jets containing a beauty quark ($b$-jets) are identified by means of a neural network (NN) built by the combination of variables describing the properties of secondary vertices and of tracks with large impact parameters relative to the primary vertex.

\subsection{Top + jet resonance (CDF)}
\begin{figure}[ht]
     \centerline{\includegraphics[width=0.467\columnwidth]{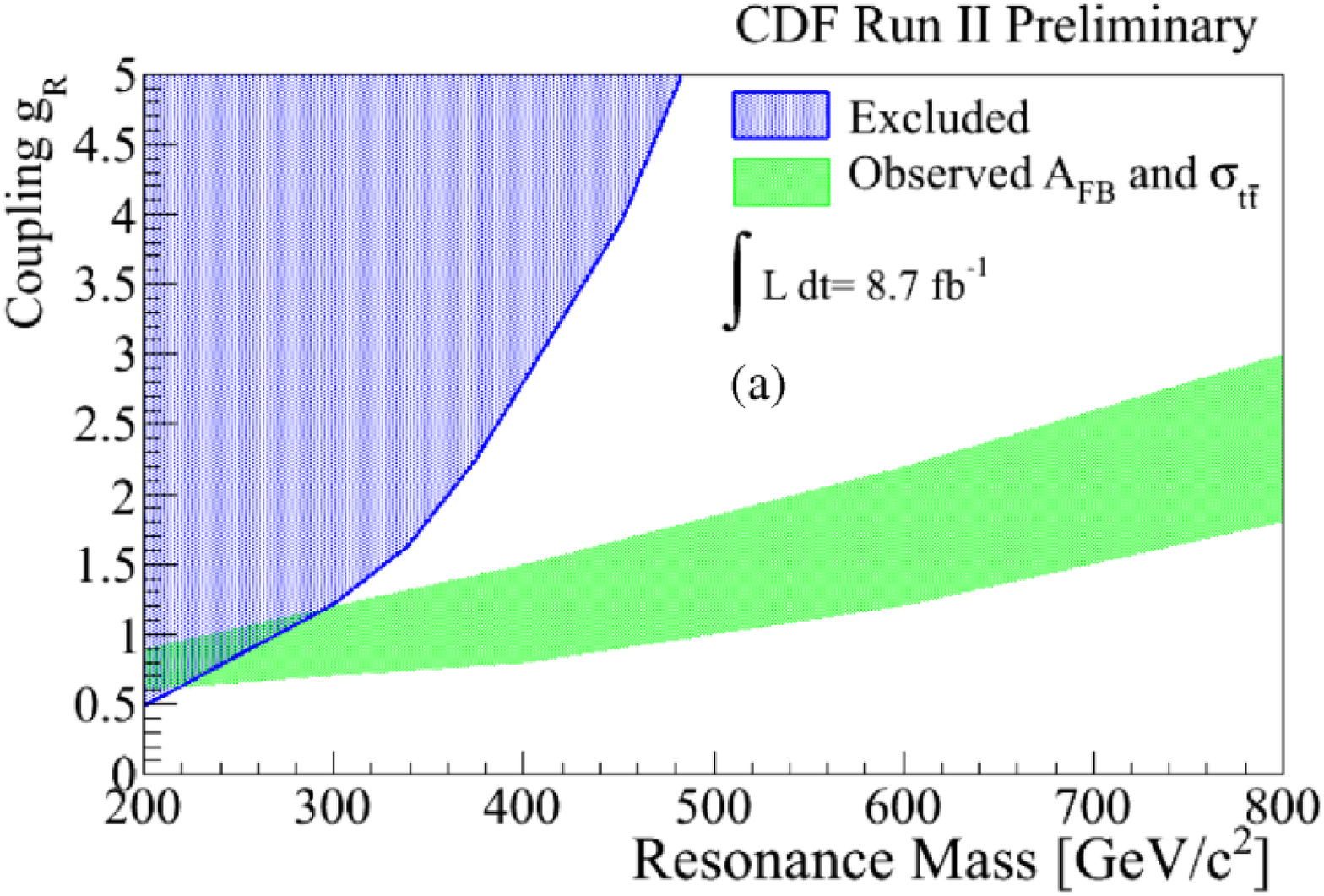}
     \includegraphics[width=0.446\columnwidth]{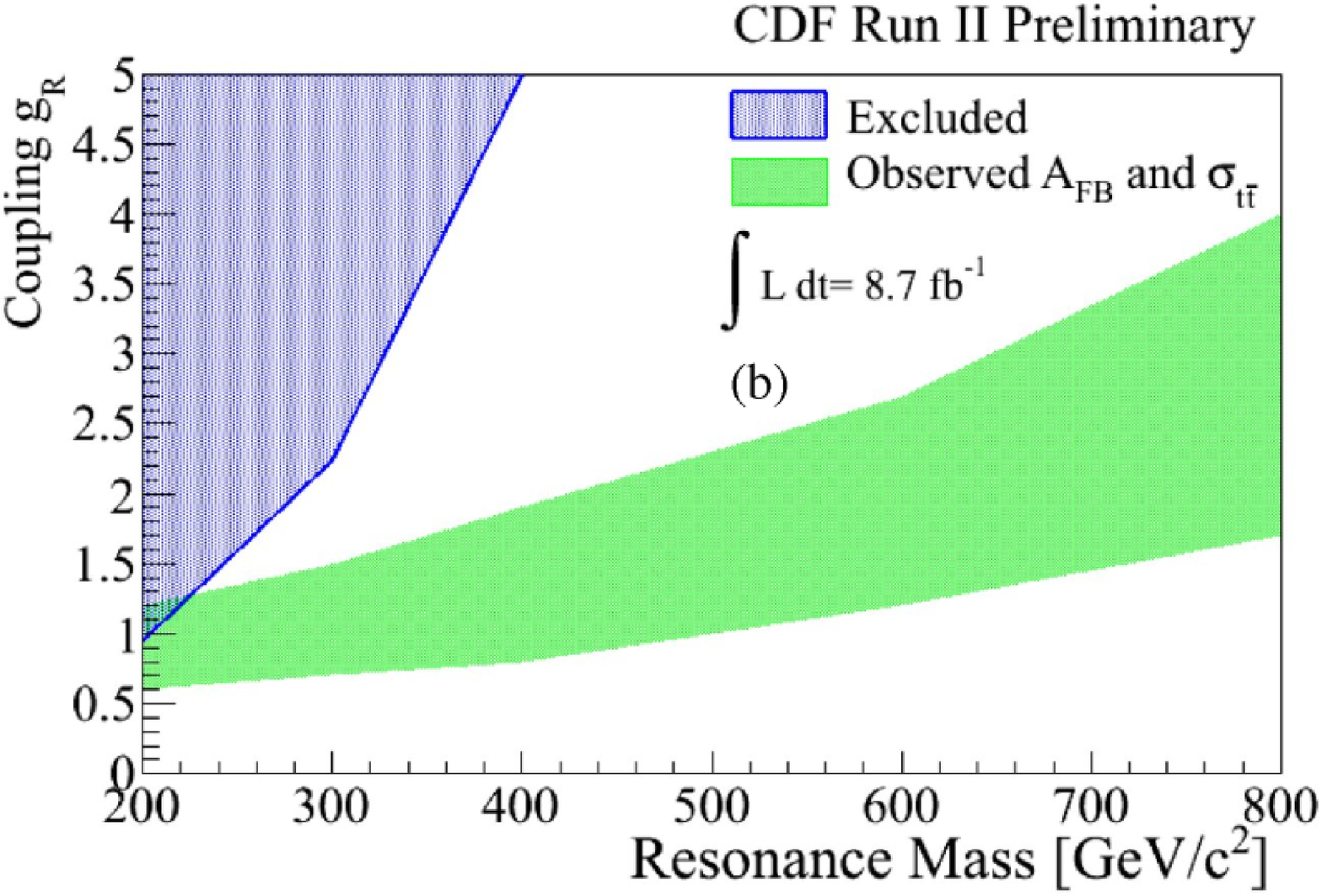}}
  \caption{\label{fig:cdfalljets_limits} (a) shows the excluded region in mass-coupling space (hashed blue) where the new heavy particle $M$ is part of a singlet or colored triplet (b). In addition the regions consistent with the observed $A_{FB}$ and \ttbar and single top cross section measurements are indicated (green band).}
 \end{figure}

A search for a heavy new particle $M$ produced in association with a top quark using $8.7~\mathrm{fb^{-1}}$ of CDF data \cite{cdfalljets} is discussed. The data sample represents the final data sample for this kind of analysis. One of the motivations of this search is the deviation of the measurement of the forward-backward asymmetry $A_{FB}$ from the SM prediction as recently reported by CDF and \dzero \cite{afbd0cdf}. The measured value of $A_{FB}$ is significantly larger than the Standard Model (SM) prediction and many models explain this by adding a new heavy particle $M$. The final state is the \ljets decay final state with five or more jets with at least one identified as $b$-jet, and missing transverse momentum \met. The resonance mass $m_{tj}$ is reconstructed by using a top kinematic reconstruction followed by a likelihood scan for the best match to the \ttbar topology. The remaining jets are paired with the $t/\bar{t}$ with $m_{tj}$ being the combination with the highest mass. 
No signal is observed and instead limits on the production of \ttbar$+ j$ via a new heavy mediator $M$ are set. Upper limits as a function of $m_{tj}$ range between $0.61$ and $0.02~\mathrm{pb}$ at 95\% confidence level (C.L.). The results have also been used to exclude two specific models in mass-coupling space. Figure \ref{fig:cdfalljets_limits} shows
the excluded regions (hashed blue) in the case that $M$ is part of a new singlet (a) or colored triplet (b). In addition the regions consistent with the observed $A_{FB}$ and \ttbar and single top cross section measurements (green band) are indicated.

\subsection{Narrow resonance (\dzero)}
\begin{figure}[ht]
     \centerline{\includegraphics[width=0.975\columnwidth]{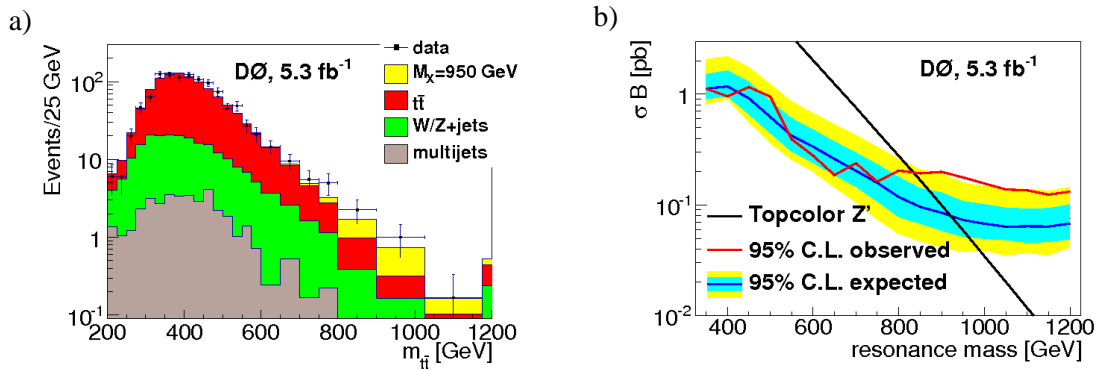}}
  \caption{\label{fig:zprimed0_limit} a) compares data from events with at least four jets to expectations from SM processes and a $950~\mathrm{GeV/c^2}$ resonance signal with the best fitted cross section times branching fraction of $\sigma \cdot BR = 0.10~\mathrm{pb}$. b) shows upper limits at 95\% C.L. on $\sigma \cdot BR$ for a narrow resonance as a function of the resonance mass. More details in the text.}
 \end{figure}

\dzero used $5.3~\mathrm{fb^{-1}}$ of data to search for a narrow resonance produced in \ttbar events \cite{narrowd0}. The final state used for the analysis is the \ljets decay final state of \ttbar events with a lepton $(e/\mu)$ and at least three additional jets with at least one of them identified as a $b$-jet, and \met. Figure \ref{fig:zprimed0_limit}a) shows the distribution of the invariant mass of the \ttbar system $m(t\bar{t})$ with at least four jets.
No observation of a narrow resonance has been made, but a slight excess of 2 standard deviations (s.d.) of events around $950~\mathrm{GeV/c^{2}}$ is seen. The best fit yields a cross section times branching fraction of $\sigma \cdot BR(M_X) = 0.10 \pm 0.05~\mathrm{pb}$. The absence of a narrow resonance allows to calculate limits for the NLO production cross section of a topcolor $Z'$ boson. The intrinsic width $\Gamma_X$ of the $Z'$ has been set to $0.012 \cdot M_X$ and a branching fraction for $Z' \rightarrow t\bar{t}$ of 100\% is assumed. Figure \ref{fig:zprimed0_limit}b) shows the upper limit on $\sigma \cdot BR$ for a narrow resonance as a function of the resonance mass. The shaded regions around the expected limit represent the one and two standard deviation bands. The solid line shows the predicted topcolor $Z'$ production cross section. The observed cross section limits exclude $Z'$ boson masses below $835~\mathrm{GeV/c^{2}}$ (95\% C.L).

\subsection{Single top + dark matter candidate (CDF)}
\begin{figure}[ht]
     \centerline{\includegraphics[width=0.467\columnwidth]{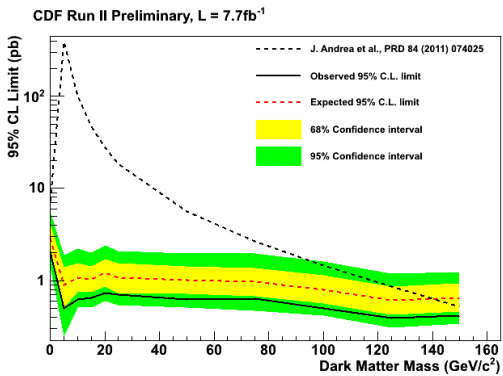}}
  \caption{\label{fig:limitDMsingletop} The figure shows the upper cross section limit at 95\% C.L. as a function of the dark matter candidate mass. More details in the text.}
 \end{figure}

A dark matter candidate can be produced in association with a single top quark. CDF used $7.7~\mathrm{fb^{-1}}$ of data to perform the first search for this specific signature \cite{singletopDM}. The final state consists of three jets with at least one identified as a $b$-jet. The dark matter signal is expected to contribute significantly at high \met. In absence of a signal upper cross section limits are calculated as a function of the mass of the dark matter candidate. Figure \ref{fig:limitDMsingletop} shows the upper cross section limit at 95\% C.L. which is about $0.5~\mathrm{pb}$ for dark matter candidate masses of $0 - 150~\mathrm{GeV/c^2}$.
The shaded regions around the expected limit represent the one and two s.d. bands. In addition the predicted dark matter candidate production cross section is indicated by the dashed black line.

\subsection{Boosted top quarks (CDF)}
CDF used $6.0~\mathrm{fb^{-1}}$ to search for a signature corresponding to boosted tops in a sample of one or two high transverse momentum massive jets with additional \met \cite{cdfboost}. The substructure of high transverse momentum objects (or jets) had not been studied extensively at Tevatron before this search. The term boosted top refers to the fact that the decay products of these top quarks are collimated into one single massive jet. The background estimation is done using data-driven methods. The predicted top cross section for $p_T > 400~\mathrm{GeV/c}$ using the MSTW2008NNLO\cite{mstw08nnlo} parton density distribution function (PDF) is $4.55_{-0.41}^{+0.50}~\mathrm{fb}$. No signal is observed and an upper cross section limit of $38~\mathrm{fb}$ at 95\% C.L. is set for $p_T > 400~\mathrm{GeV/c}$. It is also possible to search for the pair production of a massive object, in this case an upper cross section limit of $20~\mathrm{fb}$ at 95\% C.L. is derived.

\subsection{Lorentz Invariance Violation (\dzero)}
\begin{figure}[ht]
     \centerline{\includegraphics[width=0.325\columnwidth]{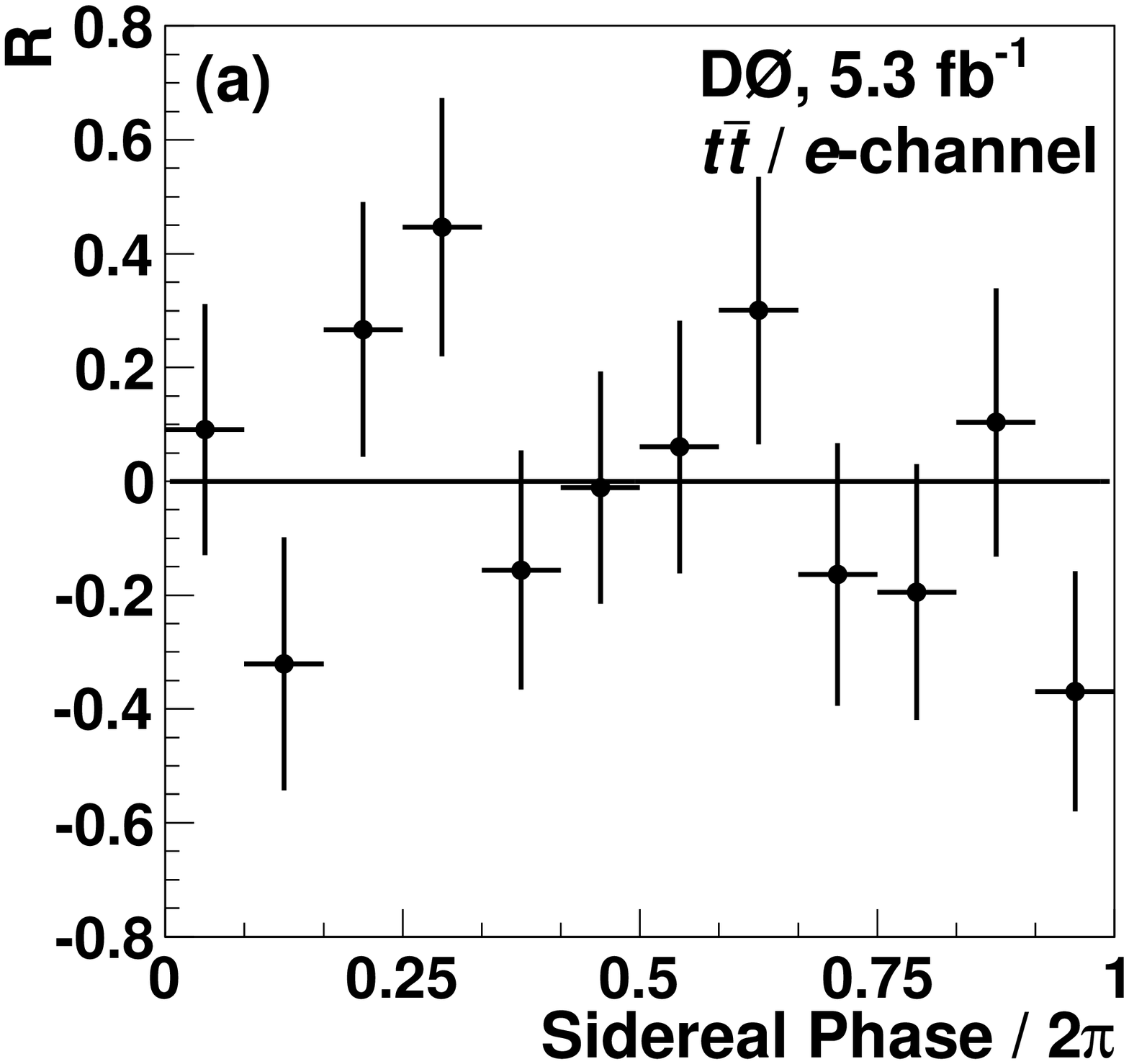}
     \includegraphics[width=0.325\columnwidth]{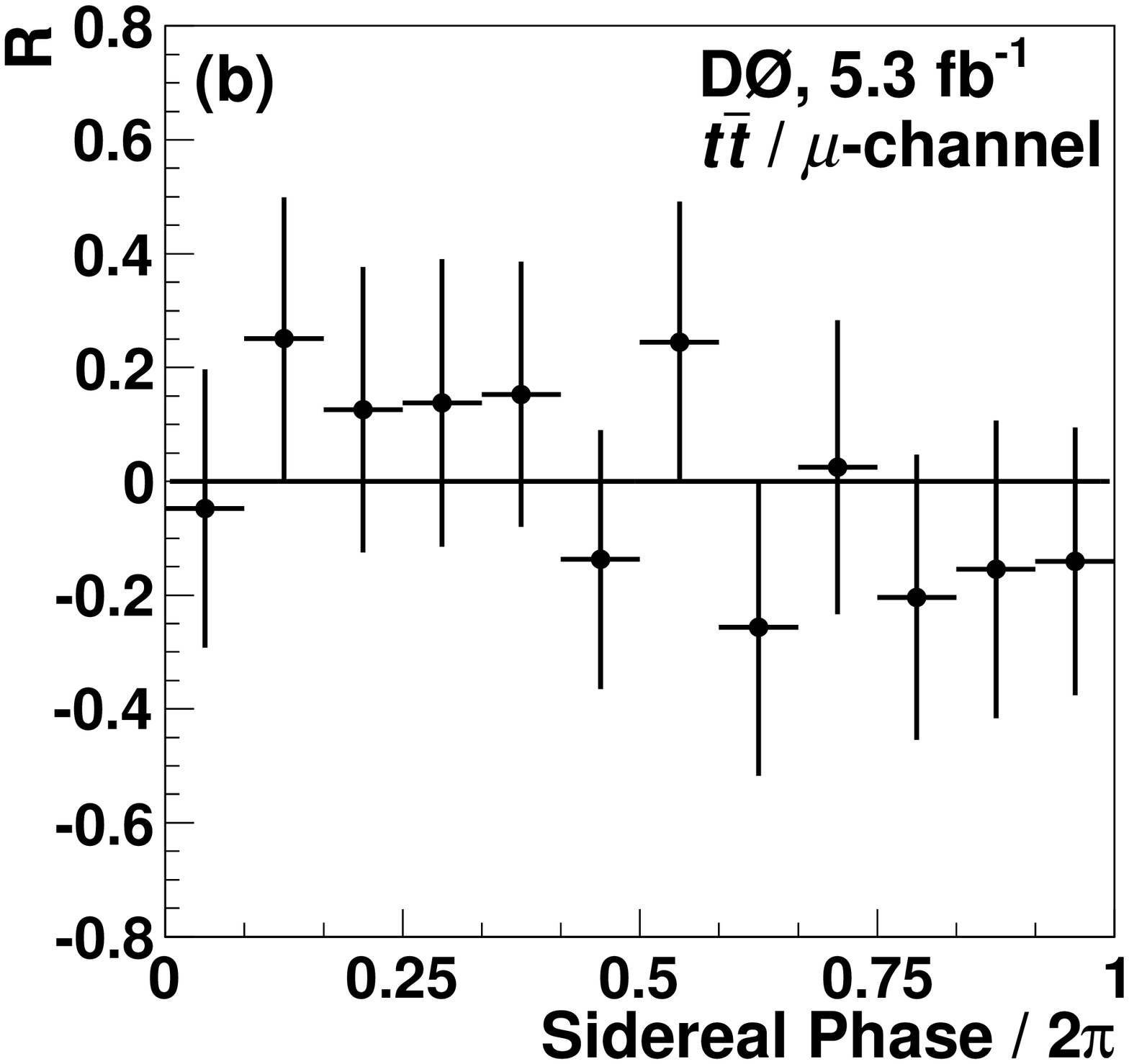}}
  \caption{\label{fig:livratio} (a) shows the background and luminosity corrected ratio $R$ as a function of the siderial phase (one siderial day) for events containing electrons, whereas (b) shows the same ratio $R$ for the muon case.}
 \end{figure}

\dzero searched for a time dependent \ttbar production cross section using $5.3~\mathrm{fb^{-1}}$ of data \cite{LIVnote}. For the analysis \ttbar events in the \ljets~final state are selected with a lepton $(e/\mu)$, additional at least four jets, exactly one jet identified as a $b$-jet and \met. In addition the analysis relies on the timestamp of the data at production time. The Standard Model Extension (SME) \cite{SMEtheory} is an effective field theory and implements terms that violate Lorentz and CPT invariance. The modified SME matrix element adds Lorentz invariance violating terms for the production and decay of \ttbar events to the Standard Model terms. The SME predicts a cross section dependency on siderial time as the orientation of the detector changes with the rotation of the earth relative to the fixed stars. The background and luminosity corrected ratio $R$ is expected to be flat within the Standard Model, i.e. no time dependency of the \ttbar production cross section. Figure \ref{fig:livratio} shows this ratio as a function of the siderial phase, \mbox{i.e.~1} corresponds to one siderial day.
There is no indication of a time dependent \ttbar production cross section. Instead this measurement sets the first constraints on Lorentz invariance violation in the top sector. As the top quark decays before it can hadronize the constraints are also the first ones for a bare quark.

\section{Conclusions}

Various recent searches in top events at the Tevatron have been discussed. More details and results are given at the \dzero and CDF webpage \cite{webpages}. There is no significant evidence for non-Standard Model signals or contributions. CDF and \dzero continue to provide unique results in the top sector and more top analyzes using the final data sample are expected to come out soon.



\section*{References}
\begin{thebibliography}{99}

\bibitem{top_disc1} F.~Abe {\it et al.} (CDF), Phys. Rev. Lett. {\bf 74}, 2626 (1995) [{\tt arXiv:hep-ex/9503002}].
\bibitem{top_disc2} S.~Abachi {\it et al.} (\dzero), Phys. Rev. Lett. {\bf 74}, 2632 (1995) [{\tt arXiv:hep-ex/9503003}].
\bibitem{cdfalljets} http://www-cdf.fnal.gov/physics/new/top/confNotes/cdf10776\_ttj.pdf (2012).
\bibitem{afbd0cdf} http://www-cdf.fnal.gov/physics/new/top/2012/LepJet\_AFB\_Winter2012/CDF10807.pdf (2012);
Phys. Rev. D 84, 112055, [{\tt arXiv:1107.4995}] (2011).
\bibitem{narrowd0} Phys. Rev. D 85, 051101, [{\tt arXiv:1111.1271}] (2012).
\bibitem{singletopDM} CDF [{\tt arXiv:1202.5653}] (2011).
\bibitem{cdfboost} http://www-cdf.fnal.gov/physics/new/top/confNotes/cdf10234\_BoostedTopConf.pdf (2011).
\bibitem{mstw08nnlo} A. D. Martin, W. J. Stirling, R. S. Thorne and G. Watt, Eur. Phys. J. C63 189-285 [{\tt arXiv:0901.0002}] (2009).
\bibitem{LIVnote} \dzero, Submitted to Phys. Rev. Lett., [{\tt arXiv:1203.6103}] (2012).
\bibitem{SMEtheory} D. Colladay and V.A. Kostelecky, Phys. Rev. D 58, 116002 (1998); V.A. Kostelecky, Phys. Rev. D 69, 105009 (2004).
\bibitem{webpages}CDF: http://www-cdf.fnal.gov/physics/new/top/top.html;\\
\dzero: http://www-d0.fnal.gov/Run2Physics/top/top\_public\_web\_pages/top\_public.html

\end{thebibliography}
\end{document}